# sRSP: GPUlarda Asimetrik Senkronizasyon İçin Yeni Ölçeklenebilir Bir Çözüm

*Ayşe Yılmazer-Metin*

İstanbul Teknik Üniversitesi,
Bilgisayar Mühendisliği Bölümü, Maslak, İstanbul
yilmazerayse@itu.edu.tr

## Öz

*Asimetrik paylaşım dinamik bir paylaşım modeli olup, paylaşılan bir verinin (**yerel**) bir paylaşan tarafından sıklıkla erişilirken, diğer (**uzak**) paylaşanlar tarafından nadir erişilmesidir. GPUlarda, özel bir destek olmaksızın, asimetrik paylaşım her erişimde ağır yüklü senkronizasyon gerektirir. Uzak senkronizasyon promosyonunun (Remote Scope Promotion, RSP) [1] tanıtımı ile yerel yaşlaşanın erişimlerine hafif senkronizasyon ile izin verilirken, ağır yüklü senkronizasyon sadece nadir gereksinim duyulan uzak erişimlerde kullanılır. RSP uzak erişimlerde yerel senkronizasyonların terfii ile veri tutarlılığını sağlar. Ne yazık ki, RSP'nin ilk gerçekleştirimi ölçeklenebilir bir çözüm değildir. Biz daha verimli ve ölçeklenebilir bir RSP gerçekleştirimi sunuyoruz. sRSP olarak adlandırdığımız bu yeni tasarım yerel paylaşanın takibine ve ağır yüklü senkronizasyon işlemlerinin seçili olarak gerçekleştirimine dayalıdır. sRSP'yi zaman detaylı Gem5-APU [2] simülatörü ile gerçekleyerek değerlendirdik ve elde ettiğimiz sonuçlar sRSP'nin 64 Hesaplama Üniteli bir GPU üzerinde performansı ortalama olarak %29 iyileştirdiğini göstermektedir.*

## 1. Giriş

Düzensiz karakteristiğe sahip uygulamaların önemli bir kısmı istenilen performanslara ulaşabilmeleri için güçlü senkronizasyon desteğine ihtiyaç duyarlar. GPUlar sıradan okuma ve yazma işlemlerini gerçekleştirirken veri tutarlılığını korumaya yönelik komplike işlemlerden kaçınırlar. Veri tutarlığını sağlamaya yönelik işlemler, seyrek kullanılacağı varsayılan senkronizasyon noktalarına kaydırılmıştır ve oldukça ağır yüklüdürler. Bu yaklaşım performansı güçlü senkronizasyon desteğine dayalı uygulamalar için pek de iyi sonuçlar sağlamaz. Kısa zaman önce senkronizasyona katılan iş-birimi sayısına bağlı senkronizasyonun performans yükünü kısıtlamaya yönelik *kapsamlı senkronizasyon* [3] tanıtılmıştır. Senkronizasyon bir grup iş-birimi arasında *yerel* olduğunda bellek sisteminde ağır yüklü önbellek işlemlerinin kullanılmasına gerek olmamakta, bu ise performansı oldukça iyileştirmektedir. Ne var ki, bu yaklaşım programlama zamanında iş-birimleri arasında iletişim ve senkronizasyonun belirli olmasını gerektirir.

Paylaşılan bir verinin sıklıkla sadece tek bir iş-birimi tarafından (*yerel-paylaşan*) kullanıldığı; diğer iş-birimlerince ise nadiren kullanıldığı (*uzak-paylaşan*) asimetrik paylaşım modeli iş-çalmaya [4] dayalı yük dengeleme gibi bazı algoritmalar için oldukça önemlidir. Asimetrik paylaşımda pek çok erişim ve senkronizasyon yerel iken dinamik olarak ortaya çıkan grup dışı erişimler genel geniş çaplı ve ağır yüklü senkronizasyonu zorunlu kılarak, yerel senkronizasyonun avantajlarından faydalanmayı olanaksızlaştırır.

Yakın zamanda tanıtılan Uzak Kapsam Promosyonu ile (RSP), çoğunluğu oluşturan yerel erişimler yerel kapsamlı senkronizasyonla desteklenirken, uzak erişimlerde yerel kapsamlı son senkronizasyonlar genel kapsama terfii ettirilerek veri tutarlılığı sağlanmış olur. Böylelikle çoğunluğu oluşturan yerel erişimlerde hafif senkronizasyon işlemlerinin getirdiği performans avantajından faydalanılırken, veri tutarlılığı korunarak uzak erişimlere de olanak sağlanmış olur. Uzak kapsam promosyonunu desteklemek amacıyla Orr v.d. [1] tarafından örnek bir gerçekleştirim gösterilmiştir. Ne yazık ki, RSPnin bu ilk gerçekleştiriminde uzak kapsam promosyon işlemlerinde, önbellek-temizleme yada önbellek veri-geçersiz-kılma işlemleri *bütün* L1 önbellekleri üzerinde uygulandığı için, RSPnin ilk gerçekleştirimi ölçeklenebilir değildir.

Biz bu çalışmada RSP için *sRSP* olarak adlandırdığımız ölçeklenebilir bir gerçekleştirim sunuyoruz. sRSP sadece seçili olarak yerel tek bir önbellek üzerinde önbellek temizleme ve önbellek veri-geçersizleme ile RSP semantiğini desteklemektedir. sRSP'de paylaşılan veri üzerinde, yerel senkronizasyon gerçekleştiren, *yerel* paylaşanı ayırt etmeye dayalı bir yöntemdir. Bu sayede ölçeklenebilir ve etkili bir yöntemle RSP çok sayıda Hesaplama Ünitesinden oluşan büyük GPUlar için ve yoğun senkronizasyon desteği gerektiren uygulamalar için de uygulanabilir bir çözüm haline gelmektedir.

## 2. GPU Çalışma Modeli ve Bellek Sistemi

Genel amaçlı bir GPU uygulaması bir veya daha fazla CPU iş-parçacığı ile binlerce GPU iş-parçacığından oluşur. Bir GPU aygıtında çalıştırılan her göreve *GPU çekirdek programcığı* adı verilir. Her bir GPU *iş-parçacığı* bir data ile eşlenerek aynı çekirdek program komutlarını paralel olarak çalıştırırlar. GPU iş-parçacığı OpenCL [5] terminolojisinde bir *iş-birimi* olarak adlandırılır. Bir GPU çekirdek programı başlatıldığında yaratılan GPU *iş-birimleri* hiyerarşik olarak organize edilirler. Ana bilgisayar iş-birimleri ile GPU iş-birimleri hep birlikte bir *sistem grubunu* oluşturur. Aynı GPU cihazı üzerinde yer alan iş-birimleri bir *cihaz grubunu* oluştururlar. Bir cihaz üzerindeki iş-birimleri *iş-gruplarına* bölünür.

Bir GPU cihazı bir takım *Hesaplama Ünitesi*nden (*HÜ*) oluşur. Bir iş-grubundaki iş-birimleri aynı HÜ üzerinde çalışırlar. Her HÜ özel bir L1 veri önbelleği içerir ve GPU cihazındaki tüm hesaplama birimleri ortak bir L2 önbelleğini paylaşır. L1 önbelleği bir iş-grubu içindeki *yerel* iletişime ve L2 Önbelleği ise, birlikte başlatılan tüm iş-grupları arasında *genel* iletişime olanak sağlar.

### 2.1. GPU'larda Bellek Tutarlılık Modelleri ve Kapsamlı Senkronizasyon Semantiği

OpenCL bellek tutarlılık modeli olarak, edinme ve yayım senkronizasyon semantiğine [6] kapsamlı senkronizasyon kavramını ekleyerek adapte etmiştir [5]. Edinme ve yayım semantiğinde paylaşılan verilerin tutarlılığının sağlanabilmesi için, iş-birimlerinin edinme ve yayım işlemleri ile senkronize olmaları gerekir. Tipik olarak bir edinme yada yayım senkronizasyon işlemi bir atomik bellek işlemi ile birleştirilir (genellikle atomik okuma ile edinme ve atomik yazma ile de yayım eşleştirilir). Edinme ve yayım senkronizasyon işlemleri diğer bellek işlemleri için bariyerler oluştururlar. Edinme ve yayım semantiğini desteklemek için, derleyici ve donanım en azından iki şeyi sağlamalıdır. (1) Edinme ve yayım senkronizasyon işlemleri etrafındaki bellek işlemlerinin sırası korunmalıdır. (2) İki iş-birimi edinme ve yayım ile senkronize olurken, donanım edinmeden sonraki okumalara yayımdan sonra gelen verinin en güncel halini sağlamak zorundadır. Bunun için yayım işleminden önceki herhangi bir güncelleme ortak bir senkronizasyon noktasına yansıtılmalı ve güncel veriler bu ortak senkronizasyon noktasından sağlanmalıdır.

Senkronizasyonun getireceği performans yükü çok sayıda iş-birimi senkronizasyona dahil olduğunda çok ağır hale gelebilir. OpenCL senkronizasyon yükünü sınırlandırmak için iş-birimlerinin hiyerarşik organizasyonuna dayalı *kapsamlı* edinme ve yayım işlemlerini tanıtır. Beş senkronizasyon kapsamı vardır: İş-birimi (wi) kapsamı, SIMD-grup kapsamı (wv) kapsamı, iş-grubu (wg) kapsamı, cihaz (cmp) kapsamı ve sistem (sys) kapsamıdır. Örneğin, iş-grubu kapsamı senkronizasyonu aynı iş-grubuna ait iş-birimleriyle sınırlar. Bu çalışmada iş-grubu kapsamlı senkronizasyon yerine *yerel/lokal* senkronizasyon, cihaz kapsamlı senkronizasyon yerine de *genel* senkronizasyon terimlerini alternatif olarak kullanılmıştır. Kapsamlı senkronizasyon ayrıca hiyerarşik bir bellek tasarımında lokal senkronizasyonu mümkün kılar. Örneğin, lokal edinme ve yayım işlemleri L1 önbellek düzeyinde sağlanabilir.

**2.2. GPU bellek Sistemi Kapsamlı Edinme ve Yayım Senkronizasyon İşlemlerinin GPU Bellek Sisteminde Tipik Gerçekleştirimi**

GPU önbellek sistemleri sıradan okuma ve yazma-birleştirme gibi basit önbellek işlemleri kullanırlar. Senkronizasyonun nadir kullanılacağı varsayılarak, bellek tutarlılığı ağır senkronizasyon işlemleriyle desteklenmeye çalışılır. Bu şekilde *genel* senkronizasyon oldukça pahalı hale gelir. Genel kapsamlı yayım işleminde, yerel bir önbellekteki tüm güncellemeler genel senkronizasyon noktasına (L2 önbelleği veya ana bellek) yansıtılmalı ve genel kapsamlı bir edinme işleminde ise yerel önbellekteki tüm bloklar atılarak, tüm okumalar genel senkronizasyon noktasından yapılmalıdır.

GPU'larda veri tutarlılığı kapsamlı edinme ve yayım işlemlerinin açık bir şekilde programcı tarafından uygun şekilde kullanılmasıyla sağlanır. *Yerel yayım* gerçekleştirimi için, senkronizasyonun parçası olan atomik işlemin L1 önbelleğinde gerçekleştirimi yeterlidir. *Genel yayım* için ise, L1 önbelleğindeki bütün kirli veri bloklarının L2 önbelleğine geri yazılması ve ilgili atomik işlem L2 önbelleğinde gerçekleşmesi gerekir. *Sistem düzeyinde yayım* işlemi için ise, önce L1 önbelleğinde, sonra L2 önbelleğinde önbellek temizleme yapılması ve ilgili atomik işlem L3 önbelleğinde veya ana bellekte gerçekleştirimi gerekir.

Güncel verilerin belirtilen bir senkronizasyon noktasından edinildiğinden emin olmak için, yerel olarak saklanmış, muhtemelen eski tüm verilerin geçersiz kılınması gerekir. Basitçe, *önbellek veri-geçersiz kılma* işlemi kullanılabilir. *Yerel kapsamlı edinim* için, yalnızca senkronizasyonun parçası olan atomik işlemin L1 önbelleğinde tamamlanması yeterli olacaktır. *Genel kapsamlı edinim* için ise, L1 önbelleğindeki tüm veriler geçersiz hale getirilmeli ve ilgili atomik işlem L2 önbelleğinde gerçekleştirilmelidir. *Sistem düzeyinde edinme* için, hem L1 hem de L2 önbellekleri geçersiz kılınmalı ve L3 önbelleğinde veya sistem belleğinde atomik işlemler gerçekleştirilmelidir.

Önbellek-temizleme işlemi için etkin bir yaklaşım Hechtman v.d. [7] tarafından sunulmuştur. Bir FIFO yapısı (Senkronizasyon-FIFO, sFIFO), GPU önbelleklerindeki tüm kirli önbellek bloklarını izlemek için kullanılır. Ne zaman yerel olarak bir önbellek bloğuna yazma yapılırsa, bu veri bloğunun adresi sFIFOya eklenir. sFIFO dolar ise, önce sFIFOdan en eski blok adresi çıkarılır ve ilgili blok bir sonraki bellek seviyeye geri yazılır. Bir önbellek-temizleme işleminde, sFIFOdaki tüm adresler FIFOdaki sıralarına göre çıkarılır ve ilgili bloklar, bellek sisteminde bir sonraki seviyeye geri yazılır. Tüm yazma işlemleri onaylandıktan sonra önbellek-temizleme işlemi tamamlanmış olur. Önbellek veri-geçersiz-kılma işlemi yapılırken ise, sFIFO yapısı kullanarak önce tüm kirli bloklarının geri yazılması gerekir ve sonra önbellekte tek çevrimli bir flaş veri-geçersiz kılma gerçekleştirilebilir.

## 3. Asimetrik Paylaşım ve Uzak Kapsam Promosyonu

Asimetrik paylaşım, olası tüm senkronizasyon katılımcılarını dahil etmek için ağır performans yüklü genel kapsamlı senkronizasyon kullanmayı gerektirir.

*Uzak Kapsam Promosyonu (RSP)* [1] ile yerel-paylaşan tarafından sık erişimlerde hafif yüklü *yerel* kapsamlı senkronizasyon kullanımına izin verilirken, ağır performans yükü uzak-paylaşan tarafından nadir kullanılan erişimlere kaydırılır. Uzak senkronizasyon gerçekleştirilirken, bir önceki yerel kapsamlı senkronizasyonun genel kapsama terfii gerekir. Bununla birlikte uzak-paylaşanın yaptığı senkronizasyonu takip eden yerel kapsamlı senkronizasyonun da genel kapsama promosyonu gerekir. RSP semantiğini uygulamak için tanıtılan üç yeni senkronizasyon komutu ve anlamları aşağıdaki gibidir:

• *Uzak edinme (rm_acq):* Uzak edinme; çalıştıran için yukarı doğru bir bariyer işlevi görür. Yerel-paylasan tarafından gerçekleştirilen son yerel kapsamlı yayım işleminin genel kapsama yükseltilmesini ve uzak-paylaşan için bir edinme gerçekleştirimini sağlar. Yerel kapsamlı yayım promosyonu, geçmiş yerel güncellemelerin genel kapsama aktarılmasını sağlar. Uzaktan edinme işlemi de, uzak-paylaşan tarafından kullanıldığında en güncel verileri genel kapsamdan çekmeyi sağlar.

• *Uzak Yayım (rm_rel):* Uzak yayım; çalıştıran için aşağı doğru bir bariyer işlevi görür. Uzak-paylaşan için genel kapsamlı bir yayım işlemi gerçekleştirir ve bir sonraki yerel kapsamda edinme işleminin genel kapsama yükseltilmesini sağlar. Yayım işlemi uzak-paylaşan tarafından yapılan tüm güncellemelerin genel kapsama aktarılmasını sağlar ve sonrasında yerel kapsamdaki edinmelerin terfii ile yerel-paylaşanın en güncel verileri genel kapsamdan çekmesini sağlar.

• *Uzak Edinme + Yayım (rm_ar):* Uzak edinme+yayım işleticisi için hem aşağı hem de yukarı yönlü bellek bariyeri işlevi görür. Son yerel-kapsamlı edinme-yayım işleminin promosyonunu sağlar ve genel kapsamlı edinme-yayım işlemlerini uzak-paylaşan için gerçekler.

RSP çalışmasında (Orr vd., 2015) yerel kapsamlı senkronizasyon işlemlerinin uzak senkronizasyon gerçekleştiriminde genel kapsama yükseltilmesi amacıyla *bütün* yerel önbelleklerde ağır yüklü önbellek-temizleme ve veri-geçersizleme işlemlerini uygulayan basit bir yaklaşım izlenmiştir ve dolayısıyla ölçeklenebilir bir yaklaşım değildir. RSP'nin daha yaygın uygulama alanı bulabilmesi için daha ölçeklenebilir bir yaklaşımla gerçekleştiriminin yapılması gerekmektedir.

## 4. Uzak Kapsam Promosyonu için Verimli ve Ölçeklenebilir Yeni Bir Yöntem: Seçili-

## temizleme ve Seçili-geçersizleme İşlemleri ile RSP Gerçekleştirimi (sRSP)

Biz bu çalışmada RSP komutlarının daha ölçeklenebilir yeni bir yaklaşımla gerçekleştirimini sunuyoruz. Bunun için paylaşılan bir veriye çoğunlukla yerel senkronizasyonla erişen yerel-paylaşanı ayırt ederek ve bu veriye uzak kapsam promosyonu ile erişimlerde gerekli önbellek-temizleme yada önbellek-veri-geçersizleme işlemlerini sadece ilgili L1 önbelleğinde yapmaya olanak verecek yeni bir yöntem sunuyoruz. sRSP olarak adlandırdığımız bu yeni yöntemde, sadece *seçilen* tek bir L1 önbelleğinde önbellek-temizleme yada önbellek-veri-geçersizleme yapıldığı için biz bu işlemleri *seçili-temizleme* ve *seçili-veri-geçersizleme* olarak adlandırıyoruz.

Sunduğumuz bu yeni yöntem sFIFO yapısını kullanarak tüm yerel kapsamdaki yayım işlemlerinin izlenmesine dayalıdır. Seçili-temizleme ve seçili-veri-geçersizleme işlemlerine sahip RSP semantiğini gerçeklemek için yerel önbelleklere iki yeni donanım yapısı ekledik ve temel GPU yazma-birleştirme önbellek protokolü üzerinde değişiklikler yaptık. Geliştirmiş olduğumuz bu yeni yöntem yine sFIFO tabanlı önbellek-temizleme ve tek çevrimli hızlı önbellek-veri-geçersizleme işlemlerini kullanmaktadır. Eklediğimiz iki yeni donanım yapısı aşağıdaki gibidir:

• *Yerel Yayım Tablosu (LR-TBL):* LR-TBL yerel yayım operasyonlarının takibi için kullanılmaktadır. Bu yapı küçük boyutta bir *CAM* (Content Addressable Memory, CAM) gibi düşünülebilir. Her yerel yayım için bir LR-TBL kaydı ve ilgili atomik işlemle ilgili bir sFIFO kaydı oluşturularak, bu iki kayıt bir gösterge ile ilişkilendirilir. Her LR-TBL kaydı bir adres (atomik işlem için belirtilen bellek adresi) ve sFIFO'da atomik işlem için oluşturulan kayda gösterge içerir. Sonra açıklayacağımız gibi, sFIFO göstergesi uzaktan-paylaşanın başlattığı seçili-önbellek-temizleme işlemlerinde bir sonlandırma işareti olarak kullanılmaktadır.

• *Terfi Almış Edinme Tablosu(PA-TBL):* Uzak kapsam promosyonun gerçekleştiriminde gereksiz önbellek veri-geçersizleme işlemlerinden kaçınmak için, yerel kapsamlı edinme terfilerine ihtiyaç duyan adreslerin kayıtları tutulmaktadır. Sadece yerel kapsamlı edinme işleminin adresi ile ilgili PA-TBL'de bir kayıt bulunduğu takdirde yerel edinme işlemi genel kapsama terfi edilmekte, dolayısıyla gereksiz önbellek-geçersizleme işlemleri engellenmektedir.

İlk RSP gerçekleştiriminde [1] olduğu gibi, sRSPde de RSP'nin üç yeni işlemi desteklenmektedir. Takip eden alt bölümlerde, sRSPnin çalışma şeklini bir örnekle açıklamaktayız. Bu örnekte, paylaşılan bir veri bir iş-grubu tarafından çoğunlukla yerel olarak erişilirken (yerel-paylaşan, wg0) ve diğer iş-grupları tarafından da nadir olarak erişilmektedir (uzak-paylaşan, wg1). Bu veri yapısı bir kritik bölüm içerisinde senkronizasyonla erişilmektedir. Yerel-paylasan HÜ0 üzerinde çalışmakta ve yerel-paylaşanın erişimleri hafif yüklü yerel-kapsamlı senkronizasyonlar ile sağlanmaktadır. Uzak-paylasan ise HÜ1 üzerinde çalışmaktadır. Uzaktan-paylasan kritik bölüme girmek için **uzak edinme (rem_acq)** ve **uzak yayım (rem_rel)** senkronizasyon işlemlerini kullanmaktadır.

### 4.1. sRSP'de Bir Yerel Yayım İşleminin Gerçekleştirimi

Bu çalışmada temel alınan GPU cihazında önbellekler sFIFO yapıları barındırır. *Şekil 1*'de gösterildiği gibi örneğimiz, paylaşılan veri (*Y*) üzerinde, wg0'in güncellemesi ile başlar. *Y*'nin blok adresi sFIFO'ya eklenir (Solda *Şekil 1*(a), solda adım ❶). Daha sonra, yerel yayım yapmak amacıyla wg0 `atomic_ST_rel_wg (L <-0)` yerel edinme işlemini gerçekleştirir. `atomic_ST_rel_wg (L <-0)` işletildiğinde, önce *L*'nin blok adresi sFIFO'ya eklenir ve sFIFO'da oluşturulan kayıt için bir gösterge (mesela *kayıt-no*) alınır (*Şekil 1*(a), sağda adım ❷). sFIFO'ya atomik yazma için eklenen kaydın ardından, LR-TBL'de *L*'nin adresiyle ilişkili bir kayıt olup olmadığına bakılır. *L*'nin adresiyle önceden bir kayıt varsa, ilgili kayıt sFIFO'dan alınan gösterge ile güncellenir. Eğer *L*'in adresi ile ilgili hiçbir kayıt yoksa, LR-TBL'da bu adres için sFIFO göstergesiyle birlikte yeni bir kayıt oluşturulur (*Şekil 1*(a), sağda adım ❸). Son olarak, `atomic_ST` yerel olarak L1 önbelleğinde işletilir ve sonuç wg0'da iş-birimine iletilir.

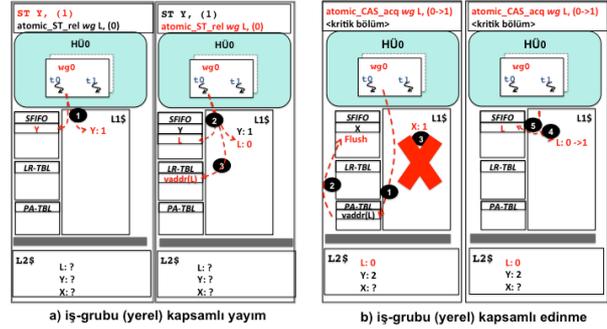

*Şekil 1. sRSP'de iş-grubu kapsamlı yayım ve edinme işlemlerinin gerçekleştirimi. (a)* wg0*'dan t1 iş-birimi Y verisini günceller. Daha sonra,* `atomic_ST_rel_wg (L <-0)` *ile yerel kapsamlı yayım işlemi gerçekleştirir. (b)* wg0*'dan bir iş-birimi* `atomic_CAS_acq_wg (L, 0->1)` *yerel kapsamlı edinme işlemini kritik bölüme girmek için işletir.*

### 4.2. sRSP'de Bir Uzak Edinme İşleminin Gerçekleştirimi

Şekil 2'de görüldüğü gibi HÜ1 üzerinde koşan wg1'den bir iş-birimi `atomic_CAS_rem_acq_cmp (L, 0 <- 1)` uzak edinme işlemini kritik bölüme girerek paylaşılan *Y* verisini değiştirmek için işletir. Uzak edinme işlemleri yerel-paylaşanın önbelleğinden (burada HÜ0'ın L1 önbelleği) paylaşılan verilerin en güncel halini çekmeyi garanti altına alırlar. Uzaktan edinme işleminin gerçekleştirimi için, ilk adımda L1_1 önbelleği (HÜ1'in L1 önbelleği) *seçili-temizleme* isteğini L2 önbelleği aracılığıyla L1 önbelleklerine gönderir (Şekil 2 adım ❷). sRSP'de bu isteğin bütün L1 önbelleklerine gönderilmesi tüm L1 önbelleklerinde temizleme işlemlerinin gerçekleştirildiği anlamına gelmez. Seçili-temizleme isteği geldiğinde, L1 önbellekleri kendi LR-TBL'lerine önceden *L*'in adresiyle kayıt edilmiş herhangi bir yerel kapsamlı yayım olup olmadığını görmek için bakarlar. Eğer *L*'in adresiyle herhangi bir yerel-kapsamlı yayım yapılmış ise bu yayımın genel kapsama terfi edilmesi gerekir. Asimetrik senkronizasyonda sadece tek bir yerel-paylaşan varsayıldığı için, tek bir L1 önbeleğinde LR-TBL kaydı bulunması beklenir. Eğer LR-TBL'de herhangi bir kayıt bulunamaz ise, hemen geri bildirim gönderilir. Yerel-paylaşan uzak edinme işlemini işleten iş-birimi ile aynı hesaplama ünitesi üzerinde çalışıyor olabilir. Bu durumda kapsam promosyonu yapmaya gerek yoktur. Çünkü yerel-paylaşan da, uzak-paylaşan da aynı L1 önbelleğini kullanmaktadırlar ve yerel paylaşan tarafından yapılan bütün güncellemeler uzak-paylaşan için de görünür haldedir. Bizim gerçeklememizde bir optimizasyon olarak öncelikle LR-TBL'de yerel bir kayıt var mı diye önce bakılarak, bulunamaz ise seçili-temizleme istekleri aynı bellek sistemindeki diğer L1 önbelleklerine gönderilmektedir.

Verdiğimiz örnekte, wg0 yerel-paylaşan olduğundan, L1_0 önbelleği üzerinde LR-TBL'de *L* adresiyle bir kayıt bulunduğu görülecektir. LR-TBL'de yer alan kayıttan, sFIFO'da temizleme yapılması gereken en son yazma işlemine bir gösterse elde edilir. Bu göstergenin işaret ettiği yazma kaydına ulaşıncaya kadar, sFIFO'dan her bir bloğun adresi çıkartılarak, ilgili önbellek bloğuna geri-yazma (*flush*) işlemi uygulanır (Şekil 2 adım ❸). Göstergenin işaret ettiği yazma (*L* adresiyle ile ilişkilendirilmiş, son yazma) işleminin kaydı da sFIFO'dan çıkartılıp geri yazması tamamlanınca, L1_0 üzerinde gerçekleştirilen seçili-temizleme işlemi tamamlanmış olur ve L1_1 önbelleğine geri bildirim yapılır (Şekil 2 adım ❹). Önbellek temizleme işlemi bittiğinde, *L*'nin adresi PA-TBL'a kaydedilir. Bu, *L* adresine gelecek bundan sonraki yerel kapsamlı edinmelerin genel kapsama yükseltilmesi gerektiğini göstermektedir. Bir başka deyişle, eğer herhangi bir yerel kapsamlı edinme *L* adresine gelirse, önbellek genel kapsam için ortak senkronizasyon noktası olan L2 önbelleğinden en güncel veriyi alabilmek için, yerel L1 önbelleğindeki tüm veriler geçersiz olarak işaretlenmelidir. L1_1 önbelleği, tüm L1 önbelleklerinden onay toplarken, kendisi de tüm kirli satırları geri yazar (yine sFIFO yardımıyla yapılır, Şekil 2 adım ❶) ve tüm önbellek veri blokları geçersiz kılar (Şekil 2 adım ❺). L1_1 üzerinde tüm veri bloklarının geçersiz olarak işaretlenmesi tamamlandıktan sonra, *uzak edinme* işlemini tamamlamak için atomic_CAS atomik işlemi L2 önbelleğine gönderilir (Şekil 2 adım ❻). L2 önbelleğinde atomik işlemin tamamlanmasıyla, sonucu wg1'den bu senkronizasyonu çalıştıran iş-birimine iletilir. Uzak-edinme işlemi gerçekleştirilirken sağlanması gereken kritik bir durum, L2 önbelleğinin *L* adresi ile ilgili veri bloğunun kitlemesi gerektiğidir. Böylece L1 önbelleklerinden bu veri bloğuna gelen herhangi bir veri okuma işlemine izin verilmez. Ayrıca, L1_1'in hesaplama ünitesinden gelen diğer istekleri uzak-edinme işlemi tamamlanıncaya kadar bekletmesi gerekir.

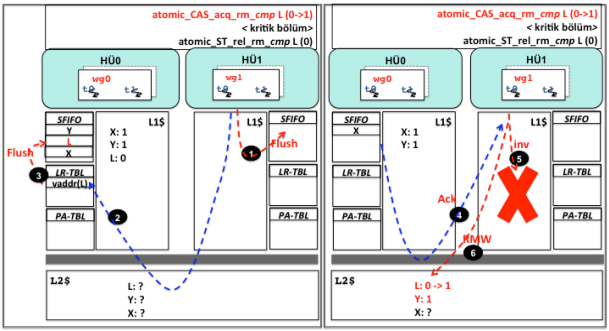

*Şekil 2. sRSP'de uzak-edinme işleminin gerçekleştirimi. HÜ1 üzerinde* wg1*'den bir iş-birimi* atomic_CAS_rem_acq_cmp (L, 0->1) *uzak edinme işlemini kritik bölüme girmek için işletmektedir.*

### 4.3. sRSP'de Bir Uzak Yayım Senkronizasyon İşleminin Gerçekleştirimi

Bir iş-birimi uzaktan edime işlemini kullanarak senkronize olduğunda, artık kritik bölüme giriş yapabilir. Kritik bölüm tamamlandığında, paylaşılan verilerdeki güncellemeleri genel kapsama taşımak için bir uzak yayım işlemi gerçekleştirilmelidir. wg1'in güncellediği bu veriye HÜ0 üzerindeki yerel-paylaşan yerel senkronizasyonla eriştiği için, wg1 tarafından gerçekleştirilen güncellemelerin alınabilmesi için, bir sonraki yerel edinmenin de genel kapsama terfi edilmesi gerekir. Şekil 3'te gösterildiği gibi, kritik bölümün ardından atomic_ST_rem_rel_cmp (L <-0) işlemi wg1 tarafından işletilir. Bu işlem paylaşılan verilerdeki tüm yerel güncellemeleri genel kapsama yansıtmak için yerel bir önbellek temizleme işlemi başlatır (Şekil 3'te adım ❶ ve ❷).

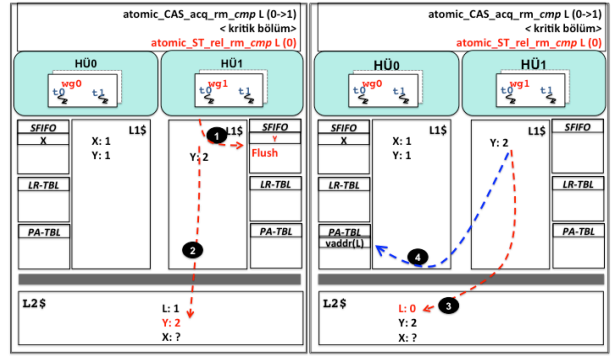

*Şekil 3. sRSP'de uzak-yayım işleminin gerçekleştirimi. HÜ1 üzerinde* wg1*'den* t1 *iş-birimi kritik bölümden çıkarken, yaptığı güncellemeleri genel kapsama yansıtmak için* atomic_ST_rem_rel_cmp (L <-0) *uzak yayım işlemini uygular.*

Yerel önbellek-temizleme işleminin ardından, L2 önbelleğinde atomic_ST işlemi gerçekleştirilir (Şekil 3'te adım ❸). Son olarak, *L* adresindeki bir sonraki yerel kapsamda edinme işleminin terfi edilebilmesi için, bir seçili-geçersizleme isteği tüm L1 önbelleklere gönderilir. *L* adresi ile seçili-geçersizleme talebini aldıktan sonra, L1 önbellekleri *L* adresini PA-TBL'ye ekler (Şekil 3'te adım ❹). Atomic_ST sonucu HÜ1'e geri döner ve uzaktan bırakma tamamlanır.

### 4.4. sRSP'de Bir Yerel Edinme İşleminin Gerçekleştirimi

Yerel kapsamda bir edinme gerçekleştirilirken, önce genel kapsama promosyonunun gerekli olup olmadığı kontrol edilmelidir. *Şekil 1* (b)'de görüldüğü gibi, örneğimizde yerel-paylaşan olan wg0, L1_0 üzerinde atomik_CAS_acq_wg (L, 1<-0) senkronizasyon komutu ile yerel kapsamda edinme işlemi başlatır. PA-TBL'de, *L* adresi ile bir kayıt olup olmadığına bakılır. Kayıt bulunur ise, L1_0 önbelleği üzerinde yerel kapsamlı edinmenin genel kapsama yükseltilerek, L2 önbelleğinden uzak kapsamla yapılan bütün güncellemelerin alınması gerektiği anlaşılır. Bu durumda önbellek veri-geçersizleme gerçekleştirilir. atomic_CAS işlemi, genel kapsamda gerçekleştirilmek üzere L2 önbelleğine gönderilir. Atomik işlem sonucu L2 önbellekten alındığında, edinme senkronizasyonu tamamlanmış olur. Önbellek veri-geçersizleme işlemi her gerçekleştirildiğinde, önbellek üzerindeki PA-TBL ve LR-TBL içindeki tüm kayıtlar temizlenecektir. Yerel kapsamda edinme sırasında PA-TBL'de bir kayıt bulunamaz ise, önbellek veri-geçersizleme işlemine gerek yoktur, atomic_CAS yerel olarak yapılarak senkronizasyon tamamlanır. atomic_CAS yazma da içerdiği için sadece *L* adresi sFIFO'ya eklenecektir.

## 5. Değerlendirme

### 5.1. Metodoloji
Bu çalışmada, OpenCL [5] ile geliştirilmiş, ve HSAIL

komutlarına [8] derlenmiş, OpenCL çekirdek programlarını çalıştırarak ve OpenCL çalışma ortamını taklit edebilen zaman detaylı Gem5 APU simülatörü [2] kullanılmıştır. Gem5 APU simülatörü, Tablo 1'de listelenen parametrelerle yapılandırılmıştır. Modellenen *temel* GPU cihazı 64 HÜ'ne sahiptir ve her HÜ dört SIMD ünitesinden oluşur. Her HÜ'nde, bir zamanlayıcı en-eski-iş-ilk algoritmasını kullanarak maksimum 40 SIMD-grubu içinden birini seçerek SIMD üniteleri üzerinde çalıştırmak için seçer. 64 iş-birimi bir SIMD-grubu oluşturur. Her HÜ bir L1 veri önbelleği içerir ve dört HÜ tarafından bir L1 komut önbelleği paylaşılır. GPU cihazı üzerinde, tüm HÜ'leri tarafından paylaşılan bir L2 önbelleği vardır. Tüm L1 veri önbellekleri ve L1 komut önbellekleri paylaşılan L2 önbelleğine bağlanır. L2 önbelleği ise sistem belleğine bağlıdır. L1 veri önbellekleri ve L2 önbelleği, sFIFO yapılarıyla donatılmıştır.

*Tablo 1. Simülasyon Parametreleri*

| L1 veri önbelleği | 16kB, 64B veri bloğu, 16-yönlü, 4 çevrim gecikmeli, 16 kayıt sFIFO |
|---|---|
| L2 önbelleği | 512kB, 64B veri bloğu, 16-yönlü, 24 çevrim gecikmeli, 24 kayıt sFIFO |
| 4 HÜ başına 1 komut önbelleği | 32kB, 64B veri bloğu, 8-yönlü, 4 çevrim gecikmeli |
| DRAM | DDR3, 8 kanallı, 500 MHz |
| L1 ve L2 Önbellek protokolü | Yer-ayırmadan, yazma-birleştirme |

Temel GPU modelimizde, genel-yayım senkronizasyon işlemleri sFIFO'yu kullanarak önbellek temizleme işlemini tetikler. Genel-edinme işlemlerinde tüm kirli veri blokları alt seviyelere yazıldıktan sonra tek çevrim önbellek veri-geçersizleme ile gerçekleştirilir. Önerdiğimiz sRSP gerçekleştirimi için, temel GPU modelinde belirtilen L1 veri önbellek ve L2 önbellek kontrolörü, yukarıda açıkladığımız seçili-geçersizleme ve seçili-temizleme işlemlerini desteklemek için değiştirilerek, geliştirilmiştir.

Değerlendirmelerimiz Pannotia [9] kıstas program grubundan seçilen *Sayfa Derecelendirme (PageRank, PRK*) ve *Tek kaynaklı en kısa yol bulma (SSSP)* çizge uygulamaları ile yapılmıştır. Bu uygulamalar kullanılan çizgeye bağlı olarak önemli yük dengesizliği yaşarlar. Bu çizge uygulamalarını, iş-çalma yaklaşımıyla dinamik yük dengeleme uygulayacak şekilde değiştirdik. Orijinal RSP çalışmasına benzer şekilde, uygulamamız Cederman ve Tsigas'in çalışmasında [10] açıklandığı gibi kilitsiz bir iş çalma uygulaması kullanmaktadır. Her bir iş kuyruğu, bir iş-grubuyla eşleştirilmiştir. Bir iş-grubunun yerel iş kuyruğu boş olduğunda, iş-grubundan iş çalmaya çalışır. Tüm iş-kuyrukları boş olduğunda çekirdek program çalışmayı sonlandırır. Yerel iş-kuyruğundan çıkartma iş kuyruğunun sonundan olurken, diğer iş-grubundan çalma o iş kuyruğunun başından olur. Böylelikle çatışmalar aza indirgenir. Kuyruktan çıkartma ve çalma sırasında senkronizasyon gerektiren çarpışma olabilir.

Çizge kıstas programları DIMACS Implementation Challenge [11] web kaynağından seçilen çeşitli giriş çizgeleri ile çalıştırılmıştır. MIS *caidaRouterLevel* çizgesi ile, PRK *cond-mat-2003* çizgesi ile ve SSSP *USA-road-BAY* çizgesi ile çalıştırılmıştır. RSPyi ilk RSP [1] çalışmasında da kullanılan beş senaryo ile değerlendirmekteyiz. Bu senaryolar aşağıdaki gibidir:

*Temel Senaryo:* Çalma işlemi devre dışı bırakılmıştır ve iş kuyruklarından iş alınırken genel kapsamlı senkronizasyon kullanılır. Temel senaryoda senkronizasyon gerekli değildir, çünkü her bir iş-kuyruğuna yalnızca kuyruk sahibi tarafından erişilir ve kuyruğa iş-ekleme yoktur. Ancak, genel kapsamlı senkronizasyon ile bu senaryo bizim için bir temel sağlar.

*Yalnızca Kapsam:* Yalnızca Kapsam senaryosu, genel kapsamlı senkronizasyondan kaçınır ve yerel kapsamlı senkronizasyon kullanır. Ancak kapsamlı senkronizasyon, dinamik paylaşım için yeterli olmadığından çalmaya izin verilmez. Performans avantajı hafif yüklü yerel senkronizasyon kullanımından gelir.

*Yalnızca Çalma:* Yalnızca çalma senaryosu genel kapsamlı senkronizasyonla çalmayı kullanır. Performans avantajı yüklerin dengelenmesinden gelir.

*RSP:* RSP'de, çoğunlukla iş-kuyruğunun sahibi tarafından yapılan erişimlerde yerel senkronizasyon; iş çalma sırasında ise uzak-edinme ve uzak-yayım kullanılır. Uzak edinme ve yayım yaparken bütün önbelleklerde temizleme ve geçersizleme yapan, RSP'nin ilk gerçekleştirimi kullanılmıştır. Performans avantajı, çoğu senkronizasyonda yerel kapsam kullanımından ve yük dengelemeden gelir.

*sRSP:* sRSP'de, çoğunlukla iş-kuyruğunun sahibi tarafından yapılan erişimlerde yerel senkronizasyon; iş çalma sırasında ise uzak-edinme ve uzak-yayım kullanılır. RSP için yeni önerdiğimiz, seçili-temizleme ve seçili-veri-geçersizleme kullanan yöntemimiz kullanılmıştır. Performans avantajı, çoğu senkronizasyonda yerel kapsam kullanımından ve yük dengelemeden gelir.

### 5.2. Değerlendirme Sonuçları

*Hızlanma Sonuçları:* Bu kısımda çalışmış olduğumuz beş senaryo için elde ettiğimiz performans sonuçları *Temel* senaryoya göreceli olarak Şekil 4'de sunulmaktadır. Sonuçlardan *Yalnız Çalma* senaryosunun PRK ve SSSP uygulamalarında fazla etkili olamadığı görülmektedir. Bunun nedeni iş-çalmaya dayalı yük dengelemenin çok fazla performans getirememesi ve getirdiği senkronizasyon yükü ile birlikte performansa katkıyı düşürmesidir. Bu nedenledir ki, bu uygulamalarda en iyi sonuç, *Kapsam* senaryosu ve sRSP ile elde edilmiştir. Maalesef RSP, HÜ sayısına bağlı iyice artan senkronizasyon yükü ile oldukça düşük performans sergilemiştir. sRSP'nin geometrik ortalamada %29 performans iyileştirmesi sergilediği görülmektedir. sRSP en iyi sonucu SSSP uygulaması için %40 olarak göstermektedir. Hatırlatmak isteriz ki, sRSP yük dengelemeden ve yerellikten gelen iyileştirmelerin katkısıyla bu sonuçlara ulaşmaktadır. RSP maalesef yük dengelemeden ve yerellikten gelen iyileştirmelerin bir kısmını belirginleşen senkronizasyon yüküyle kaybetmektedir.

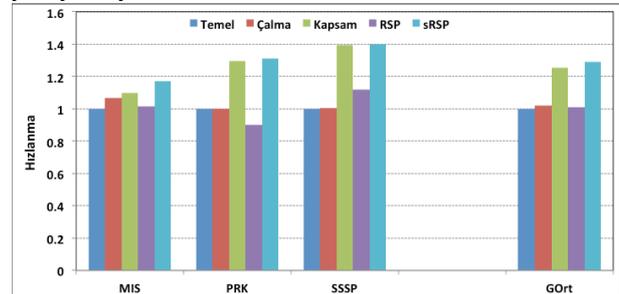

*Şekil 4. Çalışılan beş senaryo için Temel senaryoya göreceli olarak performans sonuçları.*

*Bant-genişliği Kullanımı ve Performans Yükü:* GPU'larda bellek performansını önemli ölçüde etkileyen etkenlerden biri bant-genişliği kullanımıdır. Şekil 5'da bant-genişliği kullanımında gözlemlenen iyileşme sunulmaktadır.

Bant-genişliği kullanımının göstergesi olarak, L2 önbelleğine yapılan erişimler kullanılmıştır. Şekil 6'da ise performans yükü RSP'ye göreceli olarak sunulmuştur.

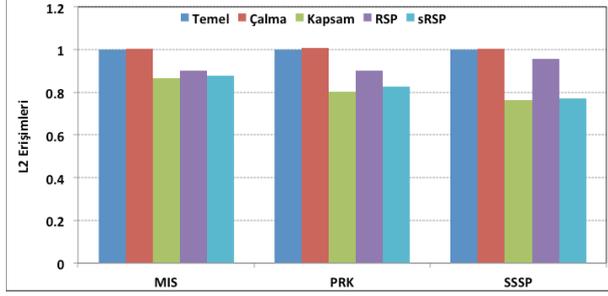

*Şekil 5. Çalışılan beş senaryo için Temel senaryoya göreceli olarak L2 erişimleri.*

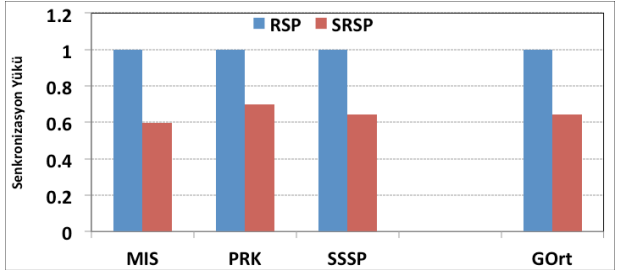

*Şekil 6. RSP ve sRPS için RSP'ye göreceli olarak performans yükü değerlendirme sonuçları.*

## 6. İlgili Çalışmalar

"Quickly Re-acquirable Locks" [12] ve "Simple and Fast Biased Locks" [13] çalışmaları asimetrik paylaşımı CPUlar üzerinde desteklemeye yönelik çözümler sunarlar. Fakat bu çözümler CPUların önbellek protokollerin sunduğu gelişmiş kabiliyetlere dayalıdır ve maalesef bu çözümleri GPUlar için kolayca genişleterek ölçeklenebilir bir çözüm elde edilmesi mümkün değildir. Bizim çalışmamıza en yakın olanları RSP [1] ve "heterogenous Lazy Release Consistency (hLRC)" [14] çalışmalarıdır. RSP önceden de belirtildiği gibi ölçeklenebilir bir RSP gerçekleştirimi sunmaz. hLRC ise ölçeklenebilirlik sağlamakla birlikte kilit yapısına dayalı olup sistem kilitlenmelerine çok açıktır. Performans her bir senkronizasyon değişkeni için ayrı blok kullanılmasını gerektirir, ve bu da önbellek verimliliğini düşürür. Ayrıca performansı önbellek ikame (replacement) politikasının uyarlanmasına bağlıdır.

## 7. Sonuçlar

Asimetrik paylaşım iş-çalma yük-dengeleme gibi uygulamalarda gerek duyulan önemli dinamik bir paylaşım modelidir. Asimetrik paylaşımı desteklemeye yönelik RSP tanıtılmış olmakla birlikte, ne yazık ki RSP'nin ilk gerçekleştirimi gereksiz yere bütün yerel önbelleklerde ağır senkronizasyon işlemleri uyguladığı için ölçeklenebilir bir çözüm değildir. Biz bu çalışmada, yerel kapsamlı senkronizasyonları takip ederek, uzak edinme ve yayım gerçekleştirirken önbellek temizleme ve önbellek veri-geçersizleme işlemlerini seçili olarak yerine getiren yeni bir yöntem tanıttık. Bu yöntemle uzak senkronizasyonun getirdiği yük azaltılarak daha iyi performans sağlanmış ve daha ölçeklenebilir bir çözüm elde edilmiştir. 64 HÜ'li büyük bir GPU cihazı için sRSP ortalama %29 performansa iyileşmesi sergilemektedir.